%% file: main.tex
\documentclass[11pt]{article}

\input{packages.tex}
\let\tmp\oddsidemargin
\let\oddsidemargin\evensidemargin
\let\evensidemargin\tmp
\reversemarginpar

\input{font_config.sty}

\input{paper_config.tex}
\usepackage{lineno}
\usepackage{tikz}
\usetikzlibrary{positioning}
\usetikzlibrary[shapes.geometric]
\usetikzlibrary{arrows.meta}

\usepackage{marvosym}

\modulolinenumbers[5]
\newcolumntype{L}[1]{>{\raggedright\let\newline\\\arraybackslash\hspace{0pt}}m{#1}}
\newcolumntype{C}[1]{>{\centering\let\newline\\\arraybackslash\hspace{0pt}}m{#1}}
\newcolumntype{R}[1]{>{\raggedleft\let\newline\\\arraybackslash\hspace{0pt}}m{#1}}

\usepackage[inline]{enumitem} 
\usepackage{verbatim}
\usetikzlibrary{mindmap,shadows}
\usepackage{tabularx}

\title{\mytitle}
\date{}

\input{authors.tex}

\begin{document}

\maketitle

\vspace{-1cm}

\begin{centering}
$^\textbf{a}$ \textit{Department of Computer Science and Artificial Intelligence, Andalusian Research Institute in Data Science and Computational Intelligence (DaSCI), University of Granada, Spain}\\
$^\textbf{b}$ \textit{Department of Software Engineering, Andalusian Research Institute in Data Science and Computational Intelligence (DaSCI), University of Granada, Spain} \\
$^\textbf{c}$ \textit{ADIA Lab, AI Maryah Island, Abu Dhabi, United Arab Emirates} \\

\end{centering}

\blfootnote{* Corresponding Author}
\blfootnote{Email addresses:  
\textbf{\texttt{dajilo@ugr.es}} (Daniel Jiménez-López),
\textbf{\texttt{rbnuria@ugr.es}} (Nuria Rodríguez-Barroso),
\textbf{\texttt{luzon@ugr.es}} (M. Victoria Luzón),
\textbf{\texttt{herrera@decsai.ugr.es}} (Francisco Herrera).}
\vspace{0.5cm}

\begin{abstract}

Deep learning models have an intrinsic privacy issue as they memorize parts of their training data, creating a privacy leakage. Membership Inference Attacks (MIA) exploit it to obtain confidential information about the data used for training, aiming to steal information. They can be repurposed as a measurement of data integrity by inferring whether it was used to train a machine learning model. While state-of-the-art attacks achieve a significant privacy leakage, their requirements are not feasible enough, hindering their role as practical tools to assess the magnitude of the privacy risk. Moreover, the most appropriate evaluation metric of MIA, the True Positive Rate at low False Positive Rate lacks interpretability. We claim that the incorporation of Few-Shot Learning techniques to the MIA field and a proper qualitative and quantitative privacy evaluation measure should deal with these issues. In this context, our proposal is twofold. We propose a Few-Shot learning based MIA, coined as the FeS-MIA model, which eases the evaluation of the privacy breach of a deep learning model by significantly reducing the number of resources required for the purpose. Furthermore, we propose an interpretable quantitative and qualitative measure of privacy, referred to as Log-MIA measure. Jointly, these proposals provide new tools to assess the privacy leakage and to ease the evaluation of the training data integrity of deep learning models, that is, to analyze the privacy breach of a deep learning model. Experiments carried out with MIA over image classification and language modeling tasks and its comparison to the state-of-the-art show that our proposals excel at reporting the privacy leakage of a deep learning model with little extra information.

\end{abstract}

\keywords{deep learning \and membership inference attacks \and data integrity \and privacy evaluation \and few shot learning}


\input{text_body.tex}

\bibliographystyle{unsrt}
\bibliography{generalbibfile}


\end{document}

%% file: packages.tex
\usepackage[dvipsnames,svgnames,table]{xcolor}
\ifdefined\directlua
  \usepackage{fontspec}
\else
  \usepackage[T1]{fontenc}
  \usepackage[nomath]{lmodern}
\fi
\usepackage{arxiv}
\usepackage{booktabs}
\usepackage{float, url}
\usepackage{subcaption}

\usepackage{nicefrac}       
\usepackage{microtype}      

\usepackage{makecell}
\usepackage{multirow}
\usepackage{stmaryrd}
\usepackage{hhline}
\usepackage{fontawesome}
\usepackage{dsfont}
\usepackage{lscape}
\usepackage{colortbl}
\usepackage{bbding}
\usepackage{amsfonts}
\usepackage{mathtools}
\usepackage{amsmath}
\usepackage{amssymb}

\usepackage{array}
\usepackage{url}
\usepackage{tablefootnote}
\usepackage{amssymb}
\usepackage{xcolor}
\usepackage{multicol}
\usepackage{todonotes, marginnote}
\usepackage{changepage,threeparttable} 
\usepackage{booktabs, multirow} 
\usepackage{soul}

\usepackage{graphicx}
\usepackage{amsmath}
\usepackage[ruled,vlined]{algorithm2e}
\usepackage{lineno}

\usepackage{todonotes, marginnote}
\usepackage{booktabs, multirow} 
\usepackage{soul}
\usepackage{tikz}
\usetikzlibrary{positioning}
\usetikzlibrary[shapes.geometric]
\usetikzlibrary{arrows.meta}
\newcommand{\OrangeCircle}[1][red!30!yellow,fill=red!30!yellow]{\tikz[baseline=-0.5ex]\draw[#1,radius=3pt] (0,0) circle ;\;}%
\newcommand{\RedCircle}[1][black!10!red,fill=black!10!red]{\tikz[baseline=-0.5ex]\draw[#1,radius=3pt] (0,0) circle;\;}%

\newcommand{\danirev}[1]{\textcolor{black}{#1}}
\newcommand{\context}[1]{\textcolor{black}{#1}}
\newcommand{\challenge}[1]{\textcolor{black}{#1}}

\newcommand{\proposal}[1]{\textcolor{black}{#1}}
\newcommand{\evaluation}[1]{\textcolor{black}{#1}}
\newcommand{\paperdesc}[1]{\textcolor{black}{#1}}
\newcommand{\nuria}[1]{\textcolor{black}{#1}}
\newcommand{\final}[1]{\textcolor{black}{#1}}

%% file: authors.tex
\author{
        Daniel Jiménez-López $^{\text{a}}$,
	Nuria Rodríguez-Barroso $^{*\text{,a}}$,
        M. Victoria Luzón $^{\text{b}}$, \\
        \textbf{Francisco Herrera $^{\text{a,c}}$}
}

%% file: text_body.tex
\section{Introduction}

\context{The increasing rate at which Artificial Intelligence (AI) systems are developed and incorporated into our routines not only has increased privacy awareness but also widened it. Data privacy is certainly a concern, as is data integrity, especially in novel settings such as federated learning \cite{RodriguezBarroso2022SurveyOF}. Moreover, the increased sense of surveillance has boosted the development of techniques that protect us from the misuse of AI systems and deepen our understanding of them \cite{camiseta}.}

\context{For a long time, deep learning models were considered black boxes, from which no information apart from what was related to the main task of the model could be extracted. This led to a false sense of security that neglects to question the privacy risks to which training data are exposed. In fact, deep learning models are known to be very dependent on their training data. Since a privacy attack can modify these data in order to change the behavior of the model and leak private information, their integrity must be protected.}

\context{There is a wide variety of privacy attacks, ranging from the ability to infer whether the training data has a certain property, to the ability to reconstruct the training data, through the ability to infer whether some data were used to train the model, namely: Property Inference Attacks \cite{PIA1}, Feature Reconstruction Attacks \cite{FIA3} and Membership Inference Attacks (MIA) \nuria{\cite{WU2023110014, manzonelli2024membership}}. Such attacks pose a great challenge to face due to the growing concern about data privacy and the prospective legal regulatory frameworks that will require safeguarding data privacy in all stages of the system that relies on AI, which is already in the currently published recommendations~\cite{gdpr2}.  \nuria{In this work, we focus on MIA \cite{li2025membership}, which pose a significant threat to the privacy of learning models by exploiting differences in how models respond to training and non-training data \cite{zhu2024fedmia}, revealing whether a specific data record was used in training, potentially exposing sensitive information.}}

\context{Fortunately, privacy attacks can be used to evaluate the privacy and integrity of training data, thus addressing the increase in awareness of data privacy and modifications in international regulations, such as the EU IA act\footnote{\url{ https://artificialintelligenceact.eu/}}. This novel viewpoint, for instance, facilitates the auditing of advanced face recognition models for identifying potential unauthorised data for training AI models using MIA\cite{Dealcala_2024_CVPR}.}

\context{To illustrate the risk of MIA, we devise the following privacy risk scenario. Let us consider a deep learning classifier with the task of determining the stage of development of a cancer type using cancer tissue collected from patients. Such a model can be attacked by employing a MIA to infer whether a person's tissue belongs to the training data of the model, concluding whether she has cancer or not. \nuria{This is a data integrity challenge as }this information should remain private, as it was provided privately to support cancer research, not to be abused by any third party. A similar privacy leakage scenario is explored in \cite{LANDAU2020105932}.}

\context{Further along this line, in early research related to MIA there was no consensus on which metric should be used to evaluate the effectiveness of MIA, which also poses an issue if we consider MIA as a tool to evaluate privacy. Recent works \cite{carlini2021membership} have also studied this problem, proposing to report the True Positive Rate at low False Positive Rates (TPR at low FPR), which is a metric commonly used in other areas related to computer science security \cite{203674, 10.1145/2808769.2808780, Lazarevic2003ACS}.}

\context{Unfortunately, TPR at low FPR has an intrinsic issue: it requires fixing a low False Positive Rate, which depends on the size of the evaluation datasets. It compares the severity of the privacy leakage not feasible between different setups, and hardens a qualitative understanding of the metric, leading to a lack of interpretability. An incorrect interpretation of privacy leakage is dangerous as it can be underestimated, exposing sensitive data, or overestimated, thus severely impacting the performance of the attacked model \cite{bagdasaryan2019differential}.}

\challenge{Nonetheless, if we consider MIA as a tool to evaluate the privacy of a deep learning model, evaluating the privacy leakage requires more computing time than training the attacked model itself. Furthermore, the evaluation demands considerable amounts of private data. Those issues are exacerbated by the increasing rate at which the complexity, computing power and data requirements of deep learning architectures grow nowadays \cite{gao2020pile, gpt3, dalle2}.}

\challenge{These problems raise the question whether MIA can perform successfully in real-world environments. The evaluation of the privacy leakage of a model should not be more resource-intensive than the training process of the victim model, nor should it require more data than that used when training the victim model. Therefore, there is a pressing need for new MIA models with \nuria{reduced} requirements as possible in terms of data availability and time/computing power consumption.}

To face these challenges we hypothesize that Few-Shot Learning techniques can be incorporated into MIA and the main evaluation metric can be reinterpreted: \begin{enumerate*}[label=(\arabic*)] \item to overcome the computational time and data availability limitations presented in the field of MIA applied to deep learning models and \item to provide an insightful view of the data integrity of a deep learning model, regardless of the MIA and victim model.\end{enumerate*}

\proposal{Thus, the contributions of this paper to the literature are twofold:}

\begin{itemize}
    \item\proposal{The incorporation of a new MIA model based on Few-Shot Learning \nuria{due to their simplicity and effectiveness}, named FeS-MIA model, to significantly reduce the number of resources \nuria{required to measure the privacy breach of a deep learning model}. It enables the assessment of the integrity of training data and devises a membership inference scenario with fewer data and computational time. We note that Few-Shot Learning tackles the problem of performing a classification task of unseen classes with as little data as possible. \nuria{Moreover, we incorporate multiple implementations of FeS-MIA models with the intent of showing the flexibility of this conceptual framework.}}
    
    \item \proposal{A privacy evaluation measure, the Log-MIA measure, that changes the scale and proportion of the reported metrics to further boost the assessment of the data integrity, leading to a reinterpretation of state-of-the-art MIA. Log-MIA is a proposal to help identify the importance of privacy leakage and raises awareness of the data integrity risk present in deep learning models. The proposed measure is not only quantitatively more reasonable, but also easier to interpret.}

    \item \proposal{Jointly, the contributions of this work provide tools to \nuria{measure} the privacy \nuria{leaks} of deep learning models, enabling to reinterpret and compare the state-of-the-art results and experimentally assess whether it is possible to achieve a significant privacy leakage with as little resources as possible.} 
    
\end{itemize}


\evaluation{To assess the proposed FeS-MIA model, we carry out extensive experiments over image classification and language modeling tasks. By resorting to the proposed Log-MIA measure, we compare FeS-MIA~model with state-of-the-art MIA methods. As a result, we confirm that almost all the MIA literature is capable of achieving a significant privacy leakage. Furthermore, we assert that it is possible to achieve a significant privacy leakage with little amounts of data and low computing resources. Altogether, we confirm with evidence that our contributions constitute a set of tools to enhance and ease the evaluation of the privacy guarantees of a deep learning model.}

\paperdesc{The rest of this paper is structured as follows: Section \ref{related.works} builds the context for our contributions. Next, in Section \ref{fs.mia}, we present our first contribution, the FeS-MIA model for more efficient attacks, followed by Section \ref{fs.eval}, which motivates and defines our second contribution, Log-MIA, an improved privacy measure. Section \ref{experimental.results} details our experimental setup and shows the performance of FeS-MIA model measured in terms of the Log-MIA measure in comparison to other state-of-the-art MIAs whose performance is reevaluated using our proposed measure. Finally, Section \ref{conclusions} concludes the paper by summarizing our findings and outlining several alternatives to expand this work in the future.}

\section{Background and Related Works} \label{related.works}

In this section, we introduce some concepts related to the contributions of this paper. Section \ref{mia.overview} introduces the state-of-the-art in the MIA field and the current metric used to evaluate them. Then, Section \ref{fs.related.works} briefly reviews the foundations of Few-Shot Learning.

\subsection{Membership Inference Attacks} \label{mia.overview}

MIA pose a data integrity risk to deep learning models. In what follows we provide a formal definition of these attacks and comment on the main issues that prevail insufficiently unaddressed in the state of the art, namely, their data and computational resources requirements, and their evaluation criteria.

Conceptually, MIA rely on a meta-classifier which, given a trained machine learning model, which we will refer to as the \textit{victim model} and some data, solves the task of detecting whether the data belongs to the training dataset of the victim model. A formal definition of MIA can be stated as follows: Let $M$ be a trained machine learning model, that is, a victim model, and $L$ a set of data, which has a non-empty intersection with $D_{M}$, the set of data used to train $M$. A MIA can be defined as a function $\phi_M: L \rightarrow \{0,1\}$ given by:
\begin{equation}
    \phi_M(x):= 
    \begin{cases}
                                   1 & \text{if }x \in D_M, \\
                                   0 & \text{if }x \not\in D_M.
    \end{cases}
\end{equation}

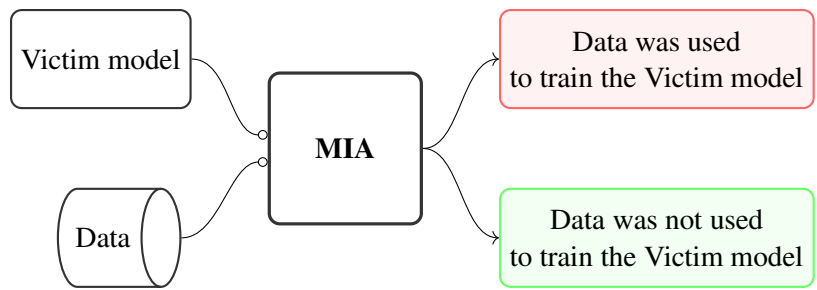
\begin{figure}[htp]
\centering
\begin{tikzpicture}[
miabox/.style={rectangle, rounded corners, draw=black!80, very thick, minimum size=20mm},
truemem/.style={rectangle, rounded corners, draw=red!60, fill=red!5,  thick, minimum size=13mm},
falsemem/.style={rectangle, rounded corners,  draw=green!60, fill=green!5,  thick, minimum size=13mm},
deepmodel/.style={rectangle, rounded corners, draw=black!80,  thick, minimum size=13mm},
data/.style={cylinder, draw=black!80,  thick, minimum size=13mm},
]
\node[miabox](mia){\textbf{MIA}};
\node[deepmodel, align=center](victimmodel)[above left=-5mm and 10mm of mia] {Victim  model};
\node[data](victimdata)[below left=-5mm and 15mm of mia] {Data};
\node[truemem, align=center](output_1)[above right=-5mm and 10mm of mia] {Data was used \\ to train the Victim model};
\node[falsemem, align=center](output_2)[below right=-5mm and 10mm of mia] {Data was not used \\ to train the Victim model};

\draw[-{Circle[open]}] (victimmodel.east) .. controls +(right:5mm) and +(left:5mm) .. (mia.170);
\draw[-{Circle[open]}] (victimdata.east) .. controls +(right:5mm) and +(left:5mm) .. (mia.190);
\draw[->] (mia.east) .. controls +(right:5mm) and +(left:5mm).. (output_1.west);
\draw[->] (mia.east) .. controls +(right:5mm) and +(left:5mm).. (output_2.west);

\end{tikzpicture}
\caption{Visual description of inputs and outputs of a MIA. It is a classifier, whose input is a machine learning model, the victim model, and some data. It outputs whether the data belongs to the training set of the victim model.}\label{fig:arch}
\end{figure}

Therefore, a MIA defined in this way is formulated as a binary classification task. Figure \ref{fig:arch} conceptually outlines this definition.

In the following, we summarize the main works of the MIA field, and comment on their requirements and evaluation procedures. 

The threat that MIA pose to deep learning models was first introduced in \cite{shokri2017membership}, revealing that it is possible to infer the membership of the training dataset of a victim model, by training binary classifiers on the output of multiple ``shadow models'' that imitate the behavior of the victim model. Such models are trained over multiple overlapping splits of similar data used to train the victim model.

An alternative approach to the development of MIA was introduced in \cite{yeom}. It exploits the difference between the average training loss and the test loss as a threshold to infer the membership of data. With the assumption that the optimal inference depends entirely on the loss function, \cite{Sablayrolles2019WhiteboxVB} derives an MIA strategy using Bayesian techniques. \cite{Jayaraman2021RevisitingMI} employs the intuition that a data point used in the training phase of the victim model must be close to a local minimum, so the loss of surrounding data points must be higher. If the data point was not present in the training dataset, similar data points must have smaller or greater loss in approximately the same proportion. Both \cite{Song2021SystematicEO} and \cite{Long2020} focus on providing good thresholds for individual samples, highlighting that some samples are more vulnerable than the others. A better calibration of the loss threshold according to the difficulty of correctly classifying the target sample is proposed in \cite{Watson2021OnTI}, rendering better performance than previous MIA. \danirev{By incorporating shadow models to estimate the train and test loss distributions and the likelihood ratio test to find the threshold with the best trade-off between precision and recall of the attack, \cite{carlini2021membership, ye2022} improve on previous works. Both propose quite similar attacks in terms of design and performance, being the main difference that \cite{ye2022} focuses on a systematic methodology to justify the design of their attacks. \cite{zarifzadeh2024lowcost} generalizes upon previous ideas employing a more sophisticated statistical test, showing that \cite{carlini2021membership} is a particular case of their proposal. It also enables then to propose the first MIA with high performances with as low as 2 shadow models, that is, they significantly reduce the computational budget required to run its attack. Lastly, \cite{bertran2024} proposes a different approach, that employs quantile regression model on a model trained minimizing the pinball loss to perform the attack, its is victim model-agnostic, given that it does not depend on the architecture of the victim model. Although it is less computationally intensive than \cite{carlini2021membership, ye2022}, it is similar in performance to them.}

\paragraph{\textit{Data and computational resources requirements}} At the time of writing this work, the state-of-the-art MIA is dominated by the attacks presented in~\cite{carlini2021membership}, \cite{ye2022} and \cite{zarifzadeh2024lowcost}. However, their proposal requires training 256 shadow models for each task at worst \cite{carlini2021membership, ye2022} and 2 shadow models at best \cite{zarifzadeh2024lowcost}, which is especially intensive for language modeling tasks. For instance, the one under study for the WikiText-103 \cite{Merity2017PointerSM} uses a GPT-2 model \cite{radford2019language} that amounts up to 1.5 billion trainable parameters. Furthermore, it requires complete availability of the dataset used to train the victim model. The assumption of availability of the training procedure and architecture of the victim model and the availability of large amounts of data are usual in the literature, but far from being feasible in practical settings.

\paragraph{\textit{Evaluation}} Carlini et al. \cite{carlini2021membership} showed that metrics such as the accuracy, precision and recall or AUC are not suitable for reporting the performance of a MIA, mainly because the binary classification task of a MIA is not a standard one. MIA solve a binary classification task where members and non-members of the training dataset of the victim model are the positive class and negative class, respectively. True Positives (TP), members of the training dataset whose membership is correctly inferred by the attack, are far more valuable than True Negatives (TN), non-members of the training dataset whose membership if correctly inferred, given that they represent data points whose membership is considered a privacy leakage. They showed that most of the current works in the field excel at detecting TN but not at detecting TP. \danirev{Moreover, a high proportion of FP can lead to a misleading perception of MIA performance \cite{reza2021}.} Thereby, metrics such as accuracy or precision are flawed. 

To solve this issue, they proposed the usage of TPR at low FPR, which is common is other computer science security areas \cite{203674, 10.1145/2808769.2808780}, and reevaluated the state-of-the-art with it. The main idea behind this metric is to inform about how many members of the training dataset are revealed (TP) when a fixed small amount of non-members are misclassified (FP). In addition, they proposed a powerful MIA that trains multiple shadow models and uses them to approximate the distribution of outputs of the loss function for members and non-members of the training dataset of the victim model. Then, they use such distributions to infer the membership of data points according to their loss value.

\subsection{Few-Shot Learning} \label{fs.related.works}

In the Few-Shot Learning field, a typical classification task is normally referred to as $N$-way $K$-shot classification, in which $N$-way stands for $N$ classes and $K$-shot represents $K$ training samples in each class. A typical $N$-way $K$-shot classification task classifies a test example into one out of $N$ unique classes based on $N\times K$ labeled training samples. 

We formally define the Few-Shot classification problem as in \cite{Dhillon2020A}. Let $(x, y)$ denote a target sample and its label, respectively. The training and test datasets are $D_s=\{(x_i,y_i)\}^{N_s}_{i=1}$ and $D_q=\{(x_i,y_i)\}^{N_q}_{i=1}$, respectively, where $y\in C_t$ for a reduced set of classes $C_t$. The training dataset is known as \emph{support set}, whereas the test dataset is known as \emph{query set}. Jointly, they are referred to as a Few-Shot episode. The number of classes, or \textit{ways}, is $N=|C_t|$. The set $\{x_i | y_i=c, (x_i, y_i) \in D_s\}$ is the support of class $c$, and its cardinality is $K$, known as \textit{shots}. The number $K$ is small in the Few-Shot setting. The set $\{x_i | y_i=c, (x_i, y_i) \in D_q\}$ is the query of class $c$ and its cardinality is $q$, known as \emph{query shots}. In the related literature $K$ is usually set equal to $1$, $5$ or $10$, whereas the Query set is fixed to 15 samples per class \cite{wang2020generalizing}. 

The goal of Few-Shot Learning is to learn a function $F$ to exploit the training set $D_s$ towards predicting the label of a test datum $x$, where $(x,y) \in D_q$, by: 
\begin{equation}
  \hat{y} = F(x; D_s).
\end{equation}

In addition to the training set, one can have a meta-training set, $D_m = \{(x_i, y_i)\}^{N_m}_{i=1}$ where $y_i \in C_m$, with the set of classes $C_m$ being disjoint from $C_t$. The goal of meta-training is to use $D_m$ towards inferring the parameters of the Few-Shot Learning model $F$. In such a case, we will denote it as $F_M$.

A Few-Shot episode is small by definition and in practice, when sampled from a larger dataset, the performance of a Few-Shot model in a single Few-Shot episode is questionable because of the randomness of the sample. To overcome this issue, Few-Shot Learning techniques are evaluated using the average result of multiple Few-Shot episodes within a small confidence interval.

A Few-Shot classification task can be approached from many angles. Based on which aspect is enhanced using prior knowledge, we can categorize the approaches into the following \cite{wang2020generalizing}:
\begin{itemize}
    \item \textit{Training data}: these methods use prior knowledge to augment $D_s$ and significantly increase the number of samples. Standard machine learning models can be used over the augmented data, and a more accurate Few-Shot model can be obtained \cite{NEURIPS2018_1714726c, 8237590}.
    \item \textit{Model}: these methods use prior knowledge to constrain the complexity of the space in which model F lies, learning a special model or embedding designed for a specific problem \cite{Ye2020FewShotLV, Liu2020PrototypeRF}.
    \item \textit{Algorithm}: these methods resort to prior knowledge to improve the best parameters of a meta-trained model, that is, to refine meta-trained models through an algorithm \cite{Dhillon2020A, Ziko2020LaplacianRF, NEURIPS2020_196f5641}.
\end{itemize}

This work focuses on Few-Shot techniques that enhance the Few-Shot algorithm by using prior knowledge, given that we will deal with meta-trained models.

\section{\nuria{Few-Shot Learning MIA model}} \label{fs.mia}


MIA are useful tools to measure the privacy of a deep learning model. However, MIA perform poorly and \danirev{are not feasible} in environments outside the research field because: \begin{enumerate*}[label=(\arabic*)] \item evaluating the privacy leakage of a model should not be more resource-intensive than training the \danirev{victim} model itself, and \item MIA should require fewer data than that used to train the \danirev{victim} model. \end{enumerate*} The application of Few-Shot techniques to the MIA field can solve these issues. However, not every Few-Shot method \danirev{for classification tasks} can be applied to create a MIA because \danirev{the task that MIA pose} is not a standard binary classification task. 

Accordingly, we enunciate the FeS-MIA model and some of its implementations, to provide a formal definition of the FeS-MIA model. 

Given a Few-Shot technique $F$, \danirev{with certain restrictions that we will specify later on}, and a victim model $M$, we define a FeS-MIA applied to $M$ as $\theta_M: L \rightarrow \{0,1\}$, with $\theta_M: = F_M$, where $L$ \danirev{has non-empty intersection with $D_M$, the dataset used to train $M$}. The training and test sets of $\theta_M$ are $D_s:=\{(x_i, y_i) | x_i \in L, y_i\in \{0,1\}\}$ and $D_q:=\{(x_i, y_i)  | x_i \in L, y_i\in \{0,1\}\}$, \danirev{that is, the support and query sets, respectively, in Few-Shot Learning terminology}. We note that classes 1 and 0 are identified with members and non-members \danirev{of $D_M$, the positive and negative classes, respectively. The sizes of the classes present in the support and query set are the cardinalities} $K=|\{x_i | y_i=c, (x_i, y_i) \in D_s\}|$ and $q=|\{x_i | y_i=c, (x_i, y_i) \in D_q\}|$ for class $c\in\{0,1\}$ are restricted to sets $\{1, 5, 10\}$ and $\{15\}$, respectively. In Few-Shot terminology, the FeS-MIA model is made of 2-way 1, 5 or 10-shots tasks.

\danirev{The main restriction for a Few-Shot technique to be employed in a FeS-MIA model is requiring a meta-trained model which, in this MIA setting, is the victim model. As is, the FeS-MIA model does not fit the usual classification of MIA based on the adversarial knowledge of the victim model: white-box or black-box MIA \cite{MIAsurvey}, given that it does not impose any restriction on the access to the victim model. It can be instantiated to be either a white-box or a black-box MIA, in both cases, it is a restriction that is imposed by the Few-Shot technique in use}.

\nuria{As we mentioned,} we instantiate our proposed model using the following Few-Shot techniques, due to their simplicity and effectiveness. \danirev{Moreover, we incorporate multiple implementations of FeS-MIA models with the intent of showing the flexibility of this conceptual framework.}

\paragraph{\textit{FeS-MIA Transductive Tuning (FeS-MIA TT)}} It employs the Transductive tuning technique proposed in \cite{Dhillon2020A}. It adds a cross-entropy classifier on top of the outputs of a meta-trained model and fine-tunes the entire model, using $D_s$ with a regularization term applied to unlabeled query samples. Specifically, the applied regularization term uses the Shannon Entropy over the predictions of the Few-Shot model on the unlabeled query set. The main idea behind this regularization term is to make the model more confident in its predictions.

\paragraph{\textit{FeS-MIA Simple-Shot (FeS-MIA SS)}} It employs the Simple-Shot technique proposed in \cite{wang2019simpleshot}. The meta-trained model is used to generate a set of normalized centroids, each representing a class of $D_s$. These centroids are used as class representatives in the logits space to classify the logits of $D_q$ using the class of the nearest centroid.

\paragraph{\textit{FeS-MIA Laplacian-Shot (FeS-MIA LS)}} It utilizes the Laplacian-Shot (LS) technique proposed in \cite{Ziko2020LaplacianRF} to improve the aforementioned SS approach, incorporating a regularization term that integrates two restrictions: (1) assigning query set samples to the nearest class centroid, and (2) pairwise Laplacian potentials, encouraging nearby query set samples to have consistent predictions.

\danirev{Given that FeS-MIA TT fine-tunes the entire model, it requires white-box access to the victim model, it is a white-box MIA. In contrast, FeS-MIA SS and FeS-MIA LS only require access to the outputs of the model, that is, both are black-box MIA.}

We remark that the above are implementations of the FeS-MIA model that showcase its strength. However, the FeS-MIA model is not tied to any particular Few-Shot technique, as long as it fits our model definition. 

\section{Towards the new privacy evaluation Log-MIA measure} \label{fs.eval}

This section is dedicated to uncovering one of the main problems we have found in the evaluation scheme usually employed for MIA. First, we show the lack of interpretability of existing metrics in Section 4.1, which serves as a motivation to reinterpret and refine them in Sections 4.2 and 4.3. Finally, the last two sections are reserved to the definition of the Log-MIA measure (Section 4.4), and the reevaluation of existing MIA under this new privacy measure (Section 4.5).

\subsection{On the lack of interpretability of TPR at low FPR} 

\danirev{The accuracy metric, commonly used in classification tasks, is not appropriate for MIA. It weights equally being a member or not of the training dataset of the victim. However, inferring correctly a member implies a higher privacy risk than inferring a non-member, moreover the cardinality of classes, member and non-member, differs significantly in size. Thus it is certainly easier to infer correctly a record not being a non-member.} The state-of-the-art MIA evaluation metric, TPR at low FPR, is a plausible solution to these problems, but it has an intrinsic issue: the interpretation of low FPR is bound to the size of the test dataset. It requires fixing FPR for each dataset, which hinders the comparison among differently sized datasets. Consequently, we are unable to find a suitable comparison of FeS-MIA model with the state-of-the-art, at least in terms of the TPR at low FPR metric. 

\danirev{We empathize that while many datasets can contain sensitive and personal information, not every combination of model and training dataset suffers the same exposure risk, requiring evaluation of MIA on multiple datasets and consequentially, diverse victim models, to properly evaluate MIA. This claim can be discussed from the perspective of an adversary that wants to test the capabilities of MIA in multiple scenarios. It has been shown in \cite{poisonMIA}, that an attacker can inject poisoned samples in the training dataset of the victim model to foster the effectiveness of MIA. Thus, the presence of outliers, either naturally occurring due to the presence of memoization \cite{Carlini2020ExtractingTD} or poisoned samples manually crafted in the training dataset, can substantially increase the success of MIA, showing that their success is related to the training dataset of the victim model.} 

\begin{table}[!htp]\centering

\scriptsize
\begin{tabular}{lrrrrrrr}\toprule
&\multicolumn{3}{c}{TPR at 0.001\% FPR} &\multicolumn{3}{c}{TPR at 0.1\% FPR} \\\cmidrule{2-7}
Method &C-10 &C-100 &WT103 &C-10 &C-100 &WT103 \\\midrule
Yeom et al. \cite{yeom} &0 \% &0 \% &0 \% &0 \% &0 \% &0 \% \\
Shokri et al. \cite{shokri2017membership} &0 \% &0 \% &- &0.3 \% &1.6 \% &- \\
Jayaraman et al. \cite{Jayaraman2021RevisitingMI} &0 \% &0 \% &- &0 \% &0 \% &- \\
Song and Mitall \cite{Song2021SystematicEO} &0 \% &0 \% &- &0.1 \% &1.4 \% &- \\
Sablayrolles et al. \cite{Sablayrolles2019WhiteboxVB} &0.1 \% &0.8 \% &0.01 \% &1.7 \% &7.4 \%  &1\% \\
Long et al. \cite{Long2020} &0 \% &0 \% &- &2.2 \% &4.7 \% &- \\
Watson et al. \cite{Watson2021OnTI} &0.1 \% &0.9 \% &0.02 \% &1.3 \% &5.4 \% &1.10\% \\
Carlini et al. \cite{carlini2021membership} &2.2 \% &11.2 \% &0.09 \% &8.4 \% &27.6 \% &1.40\% \\
\bottomrule
\end{tabular}
\caption{Comparison of MIA under the same settings for well-generalizing deep models on CIFAR-10 (C-10), CIFAR-100 (C-100), and WikiText-103 (WT103) using the True Positive Rate (TPR) at low False Positive Rate (FPR) metric. Missing values show that these MIA cannot be adapted to the natural language processing task of predicting the next word.}\label{tab:comp}
\end{table}

In Table \ref{tab:comp}, we show the TPR at low FPR values of state-of-the-art MIA on three datasets and two different tasks, image classification and language modelling. The main problem is that low FPR values shown in such table are the same in our definition of a FeS-MIA model, due to the required small test size. We have no clear arguments to justify a different FPR while keeping the ability to make meaningful comparisons of the performance of our model with the state-of-the-art.

In terms of privacy, we find that the values reported in Table \ref{tab:comp} are hard to interpret, as some values greater than zero are low and do not clarify whether their privacy leakage can be deemed negligible or not. Thus, no qualitative assessment of these results can be made. 

To clearly expose this issue, we translate some percentages presented in Table \ref{tab:comp} to raw numbers. First, when it comes to fixing a low FPR, approximately:
\begin{itemize}
    \item On CIFAR datasets, 0.001\% and 0.1\% are 0 and 25 FP, respectively. The test dataset for the MIA has 50,000 items.
    \item On the WikiText103 dataset: 0.001\% and 0.1\% are 0 and 50 FP, respectively. The test dataset for the MIA has 1,000,000 items.
    \item With FeS-MIA model, 0.001\% and 0.1\% are 0 FP because of the small test set, an intrinsic limitation of using Few-Shot techniques. We note that our test dataset has 30 items.
\end{itemize}

As shown, the significance of low FPR values increases with the size of the tested dataset. However, the size of the test dataset does not necessarily correlate with the difficulty of the task associated with it. For example, the ImageNet dataset has a test size of 150,000 elements, whereas the CIFAR datasets have 10,000 test elements. However, most deep learning models usually achieve a higher test accuracy over ImageNet than over CIFAR datasets \cite{he2016deep, efficient, wideresnets}. Hence, the meaning of low FPR must be calibrated for each dataset, as done in other fields where this metric is used \cite{203674, 10.1145/2808769.2808780, Lazarevic2003ACS, metsis2006spam}.

Secondly, when it comes to reporting the TPR at a fixed low FPR, considering only the last row of Table \ref{tab:comp}, that is, the results from Carlini et al.~\cite{carlini2021membership}:
\begin{itemize}
        \item CIFAR-10: a TPR at 0.001\% FPR equal to 2.2\% means 550 True Positives (TP) and a TPR at 0.1\% FPR equal to 8.4\% means 2,100 TP.
        \item CIFAR-100: a TRP at 0.001\% FPR equal to 11.2\% means 2,800 TP and a TPR at 0.1\% FPR equal to 27.6\% means 6,900 TP.
        \item WikiText103: a TPR at 0.001\% FPR equal to 0.09\% means 450 TP and a TPR at 0.1\% FPR equal to 1.40\% means 7,000 TP.
\end{itemize}

In conclusion: any TPR \textgreater 0 can be regarded as an accountable privacy leakage, that is, at least the true membership of one item is revealed. However, the reported TPR at low FPR might be so small that it might dangerously imply that the privacy leakage is negligible.

We have highlighted two severe issues that appear when using the TPR at low FPR metric: \begin{enumerate*}[label=(\arabic*)] \item the meaning of low FPR is ambiguous and changes for each dataset and \item TPR values can be misleading if the test dataset is too large. \end{enumerate*} This motivates rethinking how MIA are evaluated, and proposing a new privacy evaluation measure that addresses these two problems. 

\subsection{\danirev{Rethinking} low FPR} 

We fulfill the requirement of a better interpretation of low FPR, employing the natural logarithm scale. In such a scale, a change of one order of magnitude in a test dataset does not imply a drastic change in low FPR. For example, we can apply it to the datasets used in Table \ref{tab:comp}:
\begin{itemize}
    \item For CIFAR datasets, with a test dataset of 50,000 items, $\log (50,000) \approx 10$.
    \item For WikiText-103, with a test dataset of 100,000 items, $\log (100,000) \approx 11$.
    \item For FeS-MIA model, with a test dataset of 30 items, $\log (30) \approx 3$.
\end{itemize}
Therefore, a reasonably small FPR is the one achieved when rounding up to the closest integer of $\log(\text{\textit{size of test dataset}})$ FP, that is, $\lceil\log(\text{\textit{size of test dataset}})\rceil$. 

 \subsection{\danirev{A more intuitive alternative to} TPR} 
 
\begin{figure}[htp]
    \centering
    \includegraphics[width=0.8\linewidth]{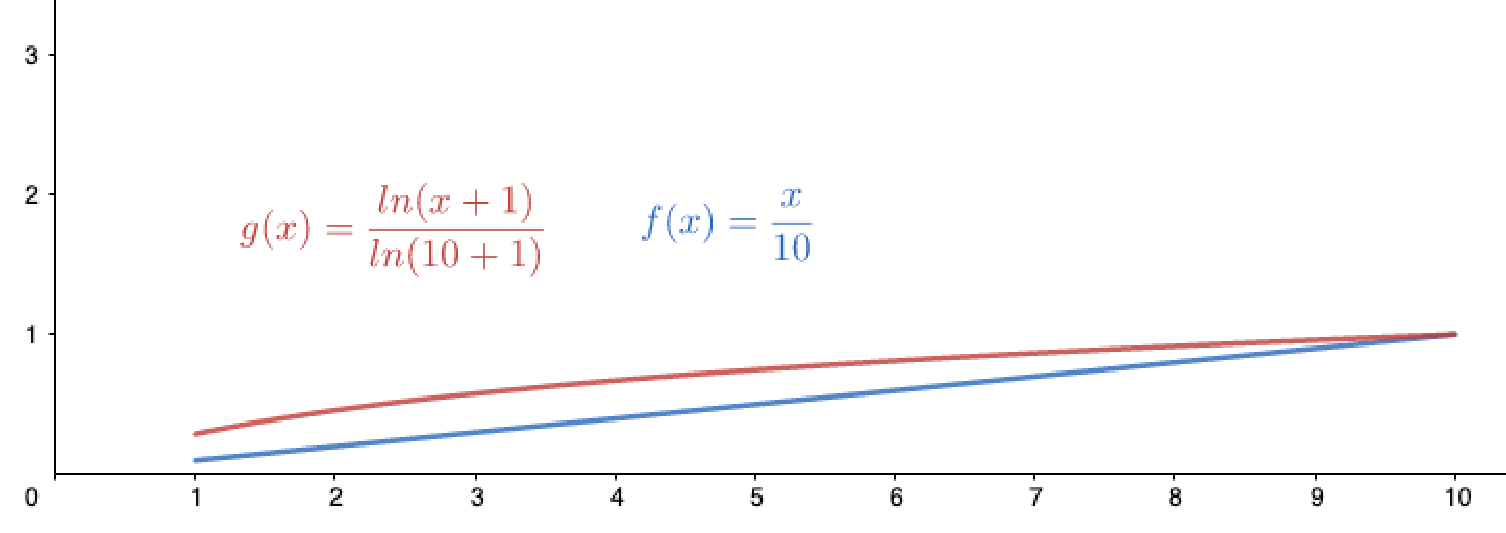}
    \caption{Comparison of growth rates of two metrics. In red the TP-log ratio and in blue the TPR with 10 members in the positive class. $x$ denotes the number of TP.}
    \label{fig:fact1}
\end{figure}

\begin{figure}[htp]
    \centering
    \includegraphics[width=0.8\linewidth]{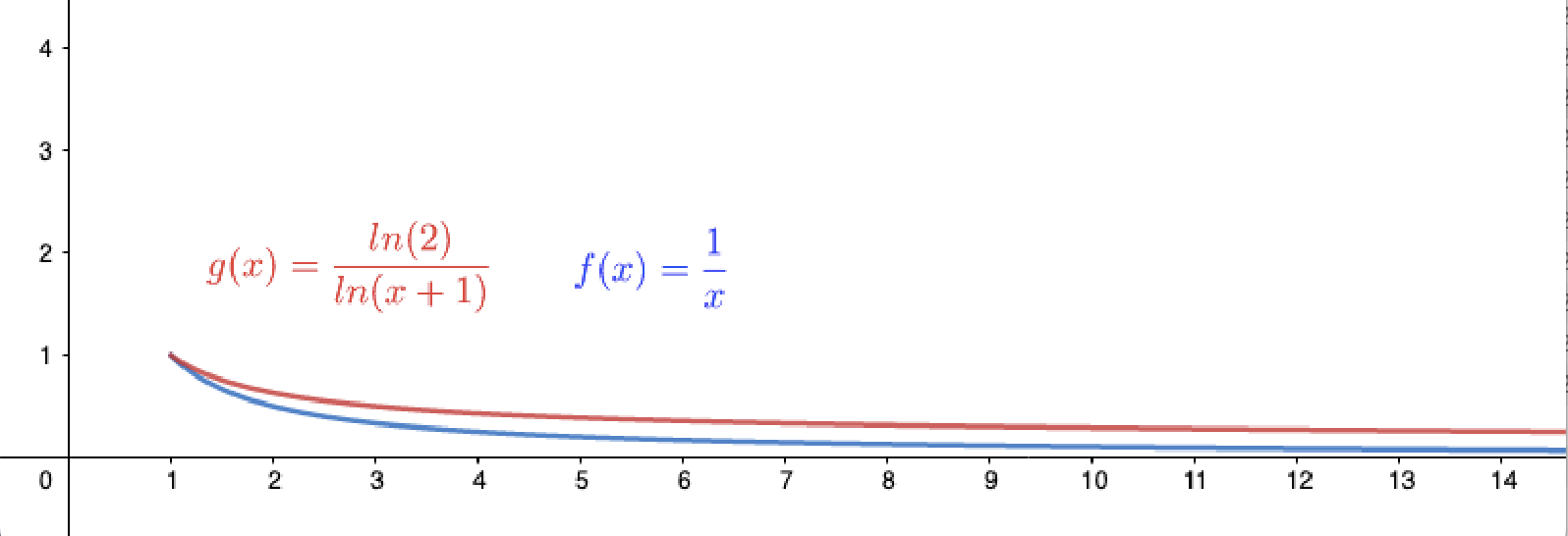}
    \caption{Comparison of the change of $\alpha$ values in both, TP-log ratio and TPR. In red and blue colors respectively, where $x$ is the size of the positive class.}
    \label{fig:fact2}
\end{figure}

 Next, we define our revision of the TPR or recall, also considering the natural logarithm scale. We propose to report the \textbf{TP log-ratio} defined as: 
\begin{equation}
    \textnormal{\textit{TP log-ratio}} = \frac{\log(\textnormal{\textit{TP}}+1)}{\log(\textnormal{\textit{size of positive class}}+1)},
\end{equation}
which has the following two interesting properties:
\begin{itemize}
    \item It grows faster than the TPR, even with small values, representing the idea that there is no negligible privacy leakage. Figure \ref{fig:fact1} illustrates the effect of this property. 
    
    \item The non-zero value closest to 0 that the \textbf{TP log-ratio} can report is $\alpha=~\frac{\log(2)}{\log(\textnormal{\textit{size of positive class}}+1)}$. That is, any value smaller than $\alpha$ means that there is no privacy leakage, whereas any value greater or equal to $\alpha$ stands for a privacy leakage. It is important to notice that this value $\alpha'$ also exists for the TPR metric, i.e., $\alpha'=\frac{1}{\textnormal{\textit{size of positive class}}}$. However, $\alpha'$ can be much smaller than $\alpha$, especially with large test sizes. This fact is supported by the inequality $ \alpha' = \frac{1}{x} < \frac{log(2)}{log(x+1)}=\alpha, \quad \forall x\ge 2 $, where x is the size of the positive class. Figure \ref{fig:fact2} shows this behavior.
\end{itemize}

\subsection{Log-MIA: a new privacy evaluation measure} 

Our interpretation of the low FPR and TPR paves the way towards a new privacy evaluation measure, Log-MIA measure, composed of two \emph{regimes}: 
\begin{itemize}
    \item \textbf{Regime A}: Report the TP log-ratio at $FP=0$. In this regime, the reported value must be greater or equal to $\alpha$ to ensure a severe privacy leakage. Otherwise, we can state that the victim model is private. 
    
    \item \textbf{Regime B}: Report the TP log-ratio at $FP = \lceil \log(\textit{size of test dataset}) \rceil$.
    In this regime we can establish further severity levels of the privacy leakage:
    \begin{enumerate}
        \item If the reported value is greater than or equal to $$\beta=\frac{log(FP+2)}{\log(\textit{size of positive class}+1)}$$ then a severe privacy leakage can be declared. The attacker can flawlessly infer the positive membership of some data used to train the victim model.
        \item If the reported value is in the interval $[\alpha, \beta)$, then there is a moderate privacy leakage. The attacker can unveil the membership of some data. However, at best it is paired with the same number of FP, that is, false memberships.
        \item Otherwise, there is no privacy leakage. The attacker cannot infer the membership of any data used to train the victim model.
    \end{enumerate}
\end{itemize}

The former is designed to address how many true memberships a MIA can infer without making any mistake, that is, its absolute values. The latter reports whether by allowing some mistakes in the positive class inferring the true membership, a MIA can still infer more true memberships than false memberships.

\subsection{Reevaluating the privacy leakage of the state-of-the-art with Log-MIA measure} 

Once the Log-MIA measure has been defined, it is possible to compare and reinterpret Table \ref{tab:comp}. Tables \ref{tab:new} and \ref{tab:new2} report the Log-MIA measure values achieved by state-of-the-art MIA. Regime A values (Table \ref{tab:new}) are computed by using the values from the TPR at 0.001\% FPR column in Table \ref{tab:comp}. For Regime B (Table \ref{tab:new2}), we cannot use directly the values provided in Table~\ref{tab:comp} due to the different FPR shown therein. Instead, we use the two ROC curve points provided to compute our required values through linear interpolation. Consequently, values shown in Table \ref{tab:new2} are computed using the expected values of the approximate ROC curve \cite{rocks}.
\begin{table}[!htp]\centering
\scriptsize
\begin{tabular}{lrrrlrlrlr}\toprule
&\multicolumn{3}{c}{TPR at 0.001 \% FPR} &\multicolumn{6}{c}{Regime A} \\\cmidrule{2-10}
Method & C-10 &C-100 &WT103 & \multicolumn{2}{c}{C-10} & \multicolumn{2}{c}{C-100} & \multicolumn{2}{c}{WT103} \\\midrule
Yeom et al. \cite{yeom} &0.0 \% &0.0 \% &0.00 \% & & 0.00 & & 0.00 & &0.00 \\
Shokri et al. \cite{shokri2017membership} &0.0 \% &0.0 \% &- & &0.00 & &0.00& &- \\
Jayaraman et al. \cite{Jayaraman2021RevisitingMI} &0.0 \% &0.0 \% &- & & 0.00 & & 0.00 & &- \\
Song and Mitall \cite{Song2021SystematicEO} &0.0 \% &0.0 \% &- & & 0.00 & & 0.00 & &- \\
 Sablayrolles et al. \cite{Sablayrolles2019WhiteboxVB} &0.1 \% &0.8 \% &0.01 \% &\RedCircle& 0.32 &\RedCircle& 0.52 & \RedCircle&0.17 \\
Long et al. \cite{Long2020} &0.0 \% &0.0 \% &- & &0.00 & &0.00 & & - \\
Watson et al. \cite{Watson2021OnTI} &0.1 \% &0.9 \% &0.02 \% &\RedCircle&0.32 &\RedCircle&0.54 &\RedCircle&0.22\\
Carlini et al. \cite{carlini2021membership} &2.2 \% &11.2 \% &0.09 \% &\RedCircle&0.62 &\RedCircle& 0.78 &\RedCircle& 0.35\\
\bottomrule
\end{tabular}
\caption{Comparison of TPR at 0.001 \%  FPR (left column) versus Regime A (TP log-ratio at 0 FP, right column). Severe privacy leakages are marked with \RedCircle, otherwise there is no privacy leakage. For CIFAR datasets (C-10 and C-100) the smallest non-zero value in the TP log-ratio is $\alpha$=0.07 and for WikiText (WT-103) it is $\alpha$=0.06. Note that in these datasets, 0.001 \% FPR is the same as 0 FP. Missing values mean that the attack cannot be applied.}\label{tab:new}
\end{table}

Focusing on Regime A (Table \ref{tab:new}), we observe that only the attacks from Sablayrolles et al. \cite{Sablayrolles2019WhiteboxVB}, Watson et al. \cite{Watson2021OnTI} and Carlini et al. \cite{carlini2021membership}, achieve a severe privacy leakage. Moreover, the MIA from Carlini et al. \cite{carlini2021membership} achieves the most severe privacy leakage. 
\begin{table}[!htp]\centering
\scriptsize
\begin{tabular}{lrrrlrlrlr}\toprule
&\multicolumn{3}{c}{TPR at 0.1 \% FPR} &\multicolumn{6}{c}{Regime B} \\\cmidrule{2-10}
Method &C-10 &C-100 &WT103 &\multicolumn{2}{c}{C-10} &\multicolumn{2}{c}{C-100} &\multicolumn{2}{c}{WT103} \\\midrule
Yeom et al. \cite{yeom} &0.0 \% &0.0 \% &0.1 \% & & 0.000 & &0.000 &\OrangeCircle&  0.206 \\
Shokri et al. \cite{shokri2017membership} &0.3 \% &1.6 \% &- &\RedCircle&0.339&\RedCircle&0.502& &- \\
Jayaraman et al. \cite{Jayaraman2021RevisitingMI} &0.0 \% &0.0 \% &- & &0.000 & &0.000 & &- \\
Song and Mitall \cite{Song2021SystematicEO} &0.1 \% &1.5 \% &- &\RedCircle&0.237& \RedCircle&0.495& &- \\
Sablayrolles et al. \cite{Sablayrolles2019WhiteboxVB} &1.7 \% &7.4 \% &1.0 \% &\RedCircle&0.516& \RedCircle&0.667& \RedCircle&0.400 \\
Long et al. \cite{Long2020} &2.2 \% &4.7 \% &- &\RedCircle& 0.533 &\RedCircle&0.608& &- \\
Watson et al. \cite{Watson2021OnTI} &1.3 \% &5.4 \% &1.1 \% &\RedCircle& 0.492 &\RedCircle& 0.643 & \RedCircle& 0.421 \\
Carlini et al. \cite{carlini2021membership} &8.4 \% &27.6 \% &1.4 \% &\RedCircle& 0.698 &\RedCircle& 0.829 &\RedCircle& 0.492 \\
\bottomrule
\end{tabular}
\caption{Comparison of TPR at 0.1 \%  FPR (left column) versus Regime B (TP log-ratio at FP = $\lceil \textnormal{log(\textit{size of test dataset}}) \rceil$. Severe privacy leakages are marked with \RedCircle, moderate privacy leakages are marked with \OrangeCircle, otherwise there is no privacy leakage. For CIFAR datasets (C-10 and C-100) the Regime B interval is [$\alpha$=0.068, $\beta$=0.245) and for WikiText (WT-103) it is
[$\alpha$=0.053, $\beta$=0.211). Missing values mean that the attack cannot be applied. }\label{tab:new2}
\end{table}

Similar conclusions can be drawn from Regime B (Table~\ref{tab:new2}), where only the attack from Yeom et al. \cite{yeom} attains a moderate leakage. From these results, we conclude that the reinterpretation of Table \ref{tab:comp}, allows reasoning that most attacks in the literature achieve a severe privacy leakage. Furthermore, the Log-MIA measure is verified to satisfy by construction the following properties:
\begin{itemize}
	\item It adapts the interpretation of low FPR to the MIA problem.
	\item It considers that any TPR\textgreater 0 is a non-negligible privacy leakage.
	\item It allows for a qualitative comparison of privacy leakages when different test sizes are used.
\end{itemize}

Thereby, we claim that Log-MIA boosts the interpretability of MIA in terms of privacy.

\section{Experimental analysis of the FeS-MIA model} \label{experimental.results}


In this section, we provide details of the experimental setup we have chosen to evaluate our proposed FeS-MIA model in subsection \ref{exp_subsec}. First, we detail which attacks, datasets and victim models are chosen, as well as, under which metrics we evaluate our results. Then, we report the results obtained and comment on them in subsections \ref{regA_subsec} and \ref{regB_subsec}.

\subsection{Experimental setup} \label{exp_subsec}

Our experiments consider multiple MIA in the literature, so that we can get a sharp picture of the performance that current MIA approaches can achieve versus our proposed FeS-MIA models under the Log-MIA measure. For the sake of clarity in subsequent discussions, we only consider MIA that attain non-zero TPR at 0.001 \% FPR (See Table~\ref{tab:comp}).

\paragraph{\danirev{\textit{Victim model and training datasets}}}
We reproduce the experimental setup employed in Table \ref{tab:comp}, that is, the datasets and deep learning models used for the victim models are\footnote{The code is available at \url{https://github.com/ari-dasci/S-few-shot-mia}}:
\begin{itemize}
    \item CIFAR-10 and CIFAR-100 \cite{cifar}: two common image classification datasets with 10 and 100 classes, respectively, approached with a Wide Resnet \cite{wideresnets} model, WRN-28-2, that is, a depth of 28 layers with a widening factor of 2. The model is trained on half of the dataset until 60\% accuracy is reached.
    \item WikiText-103 \cite{Merity2017PointerSM}: a natural language processing dataset made of articles from Wikipedia, tackled with the smallest GPT-2 \cite{solaiman2019release} model (124M parameters). The model is trained on half of the dataset for 20 epochs.
\end{itemize}
\danirev{We recall that the half used to train each model is labelled as members, and the other half is labeled as non-members. Such labels are considered when building the support and query datasets for the FeS-MIA models.}

\paragraph{\danirev{\textit{FeS-MIA model experimental setup}}}
Results of three FeS-MIA models are reported: FeS-MIA TT, FeS-MIA SS and FeS-MIA LS. The Log-MIA measure is computed over 500 runs with its 95\% confidence interval for both Regime A and B. Moreover, we consider 1-shot, 5-shots and 10-shots Few-shot Learning scenarios, with a validation set of the same size as the Query set: 15 elements per class, members and non-members of the training dataset of the victim model. The validation set permits to discover the best hyperparameter configuration for each Few-Shot Learning technique, where optimality is driven by the maximization of the privacy leakage in Regime A in each run.

We recall that FeS-MIA SS implements the Simple-Shot technique, a 1-NN classifier, so it outputs discrete probabilities, 0 and 1. Consequently, these outputs are weighted by the inverse of their distance so that the classifier can generate a continuous output in the interval $[0,1]$. We require these probabilities to compute the decision thresholds in the validation set, required for evaluating TPR at certain low FPR. Similar considerations are applied for FeS-MIA LS. We also increase the number of neighbors considered, as it provides an important boost in performance. Furthermore, for FeS-MIA SS and FeS-MIA LS, considering all the points plus the centroids of each class was found to improve significantly the privacy leakage, as will be exposed in the results. Thus, all values discussed in what follows consider these factors. 

\subsection{Regime A results and analysis}  \label{regA_subsec}

\begin{table}[!htp]\centering
\scriptsize
\begin{tabular}{llrlrlr}
\toprule
&\multicolumn{6}{c}{Regime A} \\\cmidrule{2-7}
Method &\multicolumn{2}{c}{C-10} &\multicolumn{2}{c}{C-100} &\multicolumn{2}{c}{WT103} \\
\midrule
Sablayrolles et al. \cite{Sablayrolles2019WhiteboxVB} & \RedCircle& 0.32 & \RedCircle&0.52 & \RedCircle&0.17 \\
Watson et al. \cite{Watson2021OnTI} & \RedCircle& 0.32 & \RedCircle& 0.54 &\RedCircle&0.22\\
Carlini et al. \cite{carlini2021membership} &\RedCircle&0.62 &\RedCircle&0.78&\RedCircle& 0.35\\
FeS-MIA TT 1-shot & &0.17 $\pm$ 0.02 & & 0.16 $\pm$ 0.02 & & 0.18 $\pm$ 0.02 \\
FeS-MIA TT 5-shots & &0.18 $\pm$ 0.02 & & 0.17 $\pm$ 0.02 & &  0.19 $\pm$ 0.02\\
FeS-MIA TT 10-shots & &0.18 $\pm$ 0.02 & & 0.18 $\pm$ 0.02& & 0.19 $\pm$ 0.02 \\
FeS-MIA SS 1-shot & &0.00 $\pm$ 0.00 & & 0.00 $\pm$ 0.00 & & 0.00 $\pm$ 0.00 \\
FeS-MIA SS 5-shots & &0.18 $\pm$ 0.02 & & 0.17 $\pm$ 0.02 & & 0.17 $\pm$ 0.02\\
FeS-MIA SS 10-shots & & 0.18 $\pm$ 0.02 & & 0.18 $\pm$ 0.02 & & 0.18 $\pm$ 0.02\\
FeS-MIA LS 1-shot & &0.13 $\pm$ 0.02 & & 0.12 $\pm$ 0.02 & & 0.00 $\pm$ 0.00 \\
FeS-MIA LS 5-shots & &0.15 $\pm$ 0.02 & & 0.14 $\pm$ 0.02 & & 0.00 $\pm$ 0.00 \\
FeS-MIA LS 10-shots & &0.16 $\pm$ 0.03 & & 0.14 $\pm$ 0.03& & 0.00 $\pm$ 0.00 \\
\bottomrule
\end{tabular}
\caption{Log-MIA measure, Regime A. Severe privacy leakages are marked with \RedCircle, otherwise there is no privacy leakage. For our attacks, the smallest non-zero value in the TP log-ratio is $\alpha$=0.25 for all datasets. For other attacks, on CIFAR datasets (C-10 and C-100) it is $\alpha$=0.07 and on WikiText (WT-103) it is $\alpha$=0.06.}\label{tab:regA}
\end{table}

We begin by analyzing the results obtained in the Regime A of the Log-MIA measure, shown in Table \ref{tab:regA}. The table reveals that none of the proposed set of MIA achieves a severe privacy leakage. Furthermore, it is remarkable that our values on every scenario are non-zero, and it is possible to observe that when more data is available, the scores slightly improve. Still, our values are under $\alpha$, so we cannot report any privacy leakage that is statistically significant. This situation illustrates that MIA are not easy tasks to handle with scarce data, that is, our models are not able to achieve TP with zero FP. Our results are in line with other MIA in the literature which, albeit recently published \cite{Jayaraman2021RevisitingMI, Song2021SystematicEO, Long2020}, cannot accomplish a significant privacy leakage in this regime (See Table \ref{tab:new}).

\subsection{Regime B results and analysis}  \label{regB_subsec}

\begin{table}[!htp]\centering
\scriptsize
\begin{tabular}{llrlrlr}
\toprule
&\multicolumn{6}{c}{Regime B} \\\cmidrule{2-7}
Method &\multicolumn{2}{c}{C-10} &\multicolumn{2}{c}{C-100} &\multicolumn{2}{c}{WT103} \\
\cmidrule{1-7}
Sablayrolles et al. \cite{Sablayrolles2019WhiteboxVB} &\RedCircle& 0.516& \RedCircle&0.667& \RedCircle&0.400 \\
Watson et al. \cite{Watson2021OnTI} &\RedCircle&0.492 &\RedCircle&0.643 & \RedCircle&0.421 \\
Carlini et al. \cite{carlini2021membership} &\RedCircle&0.698 &\RedCircle&0.829 &\RedCircle&0.492 \\
FeS-MIA TT 1-shot & \RedCircle&0.59 $\pm$ 0.01 & \RedCircle&0.58 $\pm$ 0.02 & \RedCircle&0.58 $\pm$ 0.02 \\
FeS-MIA TT 5-shots &\RedCircle&0.60 $\pm$ 0.01& \RedCircle&0.59 $\pm$ 0.02& \RedCircle& 0.59 $\pm$ 0.02 \\
FeS-MIA TT 10-shots &\RedCircle&0.59 $\pm$ 0.02&\RedCircle&0.58 $\pm$ 0.02& \RedCircle& 0.60 $\pm$ 0.02  \\
FeS-MIA SS 1-shot & &0.00 $\pm$ 0.00 & &0.00 $\pm$ 0.00 & &0.00 $\pm$ 0.00 \\
FeS-MIA SS 5-shots &\RedCircle&0.59 $\pm$ 0.02 &\RedCircle& 0.58 $\pm$ 0.02& \OrangeCircle& 0.57 $\pm$ 0.02\\
FeS-MIA SS 10-shots &\RedCircle&0.60 $\pm$ 0.02&\RedCircle&0.59 $\pm$ 0.02&\RedCircle &0.58 $\pm$ 0.02\\
FeS-MIA LS 1-shot &\OrangeCircle& 0.44 $\pm$ 0.03 &\OrangeCircle&0.44 $\pm$ 0.03& & 0.00 $\pm$ 0.00\\
FeS-MIA LS 5-shots &\OrangeCircle& 0.52 $\pm$ 0.02 &\OrangeCircle& 0.47 $\pm$ 0.02 & & 0.00 $\pm$ 0.00 \\
FeS-MIA LS 10-shots &\OrangeCircle& 0.54 $\pm$ 0.02 &\OrangeCircle& 0.45 $\pm$ 0.03 & & 0.00 $\pm$ 0.00 \\
\bottomrule
\end{tabular}
\caption{Log-MIA measure, Regime B. Severe privacy leakages are marked with \RedCircle, moderate privacy leakages are marked with \OrangeCircle, otherwise there is no privacy leakage. For our attacks, \textit{Regime B interval} is [$\alpha$=0.25, $\beta$=0.58) for all datasets. For other attacks the interval on CIFAR datasets (C-10 and C-100) it is [$\alpha$=0.068, $\beta$=0.245) and on WikiText (WT-103) it is
[$\alpha$=0.053, $\beta$=0.211).}\label{tab:regB}
\end{table}

We now proceed by inspecting the results of the experiments under regime B of the Log-MIA measure, which are given in Table \ref{tab:regB}. These results prove that FeS-MIA LS achieves a moderate privacy leakage in almost every Few-Shot scenario, that is, the number of TP unveiled by the attack does not overcome the number of FP. This situation requires a special consideration, as it does not expose a complete vulnerability of the victim model. However, the attacker still gains some information that might be used with additional external information to turn a moderate privacy leakage into a severe privacy leakage. It is specially feasible given the low data requirement of our proposed attacks. Accordingly, we argue that a moderate privacy leakage should not be overlooked, as it can potentially lead to a larger privacy leakage, specially with FeS-MIA.

While MIA from \cite{Sablayrolles2019WhiteboxVB, Watson2021OnTI, carlini2021membership} achieve a consistent privacy leakage in both regimes A and B, our MIA based on Few-Shot Learning elicits a different behavior in Regime B (see Table~\ref{tab:regB}) than in Regime A (see Table \ref{tab:regA}). In Regime B most of the proposed FeS-MIA models achieve a significant privacy leakage regardless of the dataset. FeS-MIA SS and Few-MIA TT produce a stable and significant privacy leakage, improving the results obtained in every scenario in the Regime A (see Table \ref{tab:regA}). That is, all the proposed models are only capable of inferring successfully the membership of some data, TP, when a small amount of FP is allowed. We note that, in most scenarios, the number of TP is greater than FP, achieving a significant privacy leakage. This behavior aligns with other attacks in the literature \cite{shokri2017membership, Song2021SystematicEO} (See Table \ref{tab:new2}).

Moreover, while LS is a technique presented in the Few-Shot Learning field as an improvement over SS, when it comes to MIA, FeS-MIA LS does not perform better than FeS-MIA SS in both regimes. This is probably because the regularization terms that LS adds on top of SS k-NN do not seem to be conceived for MIA. Still, both of them fail to achieve a significant privacy leakage in the 1-shot scenarios, while the same does not hold for TT.


In light of the previous discussion, we can conclude that in Regime B and when using the adequate Few-Shot algorithm, as few as one element per class is required to reach a significant privacy leakage, being FeS-MIA SS and FeS-MIA TT the best performing FeS-MIA models. Therefore, the FeS-MIA model is more efficient at reporting privacy vulnerabilities than the state of the art. 

FeS-MIA TT is the best FeS-MIA model overall, since it fine-tunes the entire victim model, as opposite to FeS-MIA SS and FeS-MIA LS that operate on the outputs of the model itself. We underscore that the success of our attack contribution is amplified by the fact that we require minimal amounts of data and computing resources to report a significant privacy leakage.







\section{Conclusions} \label{conclusions}

\nuria{MIA  pose a significant threat to the privacy of learning models by exploiting differences in how models respond to training and non-training data, revealing whether a specific data record was used in training, potentially exposing sensitive information.} \nuria{Detecting privacy leackage is crucial in safeguarding sensitive information from unauthorized access and misuse.} \nuria{Hence, our contributions provide a handy set of tools to interpret how private is a deep learning model, in terms of revealing the ownership of its training data as follows:}

\begin{itemize}
    \item \final{The FeS-MIA model proposes a new set of MIA based on Few-Shot Learning techniques that significantly reduces the resources required to evaluate the data integrity of a deep learning model. The techniques make the assessment of the training data integrity more feasible by requiring fewer data and computational time, matching the constraints of more realistic membership inference scenarios. }

    \item \final{The Log-MIA measure further boosts the interpretability of MIA privacy risks, leading to a reinterpretation of state-of-the-art MIA metrics. The proposed metric verifies that almost all MIA are capable of achieving a significant privacy leakage. } 
\end{itemize}

\nuria{By proactively detecting privacy leakage, this work, with the FeS-MIA model and Log-MIA measure proposals, could be crucial for enhancing the security of machine learning models, especially in scenarios where data privacy is paramount.}


\section{Acknowledgments}

\nuria{This research results from the Strategic Project IAFER-Cib (C074/23), as a result of the collaboration agreement signed between the National Institute of Cybersecurity (INCIBE) and the University of Granada. This initiative is carried out within the framework of the Recovery, Transformation and Resilience Plan funds, financed by the European Union (Next Generation).}
